\documentclass[10pt,prb,aps,twocolumn,superscriptaddress,showpacs,amsfonts,amsmath,amssymb,showkeys,floatfix]{revtex4-1}
\usepackage{bm}
\usepackage{pstricks}
\usepackage{setspace}
\usepackage{graphicx}

\begin{document}
\title{Nature of the spin-glass phase in dense packings of Ising dipoles with random anisotropy axes}
\date{\today}
\author{Juan J. Alonso}
\email[e-mail address: ] {jjalonso@uma.es}
\affiliation{F\'{\i}sica Aplicada I, Universidad de M\'alaga, 29071 M\'alaga, Spain}
\affiliation{Instituto Carlos I de F\'{\i}sica Te\'orica y Computacional,  Universidad de Granada, 18071 Granada, Spain}
\author{B. All\'es}
\email[E-mail address: ] {alles@pi.infn.it}
\affiliation{INFN--Sezione di Pisa, Largo Pontecorvo 3, 56127 Pisa, Italy}

\date{\today}

\begin{abstract}
  Using tempered Monte Carlo simulations, we study the  the spin-glass phase of  dense packings of Ising
  dipoles pointing along random axes. We consider systems of $L^{3}$ dipoles (a) placed on the sites of a simple
  cubic lattice with lattice constant $d$, (b) placed at the center of randomly closed packed spheres
  of diameter $d$ that occupy a 64\% of the volume. For both cases we find an equilibrium spin-glass phase below a temperature
  $T_{sg}$. We compute the spin-glass overlap parameter $q$ and their associated correlation length $\xi_{L}$.  From the variation of $\xi_{L}$
  with $T$ and $L$ we determine $T_{sg}$ for both systems. In the spin-glass phase, we find (a) $\langle q^{2}\rangle$ decreases algebraically with $L$, and (b) $\xi_{L}/L$ does not diverge as $L$ increases. At very low temperatures we find comb-like distributions of  $q$ that are sample-dependent. We find that the fraction of samples with cross-overlap spikes higher than a certain value as well as the average width of the spikes are
  size independent quantities. All these results are consistent with a quasi-long-range order in the spin-glass phase, as found previously
  for very diluted dipolar systems.
\end{abstract}

\maketitle

\section{INTRODUCTION}
\label{intro}

The collective behavior of systems of interacting dipoles (SID) has received renewed attention in the last few years.\cite{fiorani} This is due to
the fact that recent advances in nanoscience\cite{nano} allow to create new magnetic materials made of ensembles of identical {interacting}
nanoparticles (NP),\cite{np, bedanta} in contrast with conventional magnets. Materials built with ensembles of NP are of
interest for data storage\cite{storage} and have applications in biomedicine.\cite{bader}

Ferromagnetic NP with sizes up to a few tens of nanometers include a single magnetic domain. This domain behaves as a dipole with
magnetic moment ranging from $10^{2}$ to $10^{4}\mu_{B}$ ($\mu_B$ is the Bohr magneton). NP exhibit effective anisotropies as a consequence of
either magnetocrystalline or shape or surface effects. These anisotropies provoke the appearance of one or more easy axes in each NP, along one of which
the related dipole tends to align. Thus, the dipole is forced to overcome a certain energy barrier $E_{a}$ during any possible flipping process between
the two directions of magnetization along {one of the easy axes}.

When NP ensembles are packed in frozen arrays of well separated (not touching) particles, the dipolar becomes the only relevant
particle-particle interaction among individual NP. Moreover, for sufficiently concentrated ensembles, dipolar interparticle 
energies $E_{dd}$ may be comparable or even larger than the local $E_{a}$. In such cases, cooperative behavior among
dipoles could be observed at low temperatures\cite{bedanta} (instead of the super-paramagnetism observed 
in weakly interacting systems with $E_{dd}\ll E_{a}$).\cite{superpara}

The anisotropy of the dipolar interaction leads to geometric {\it frustration}\cite{bookmezard} when dipoles are placed in
crystalline arrays. This results in collinear antiferromagnetic (ferromagnetic) order in simple cubic lattices (face
and body centered lattices) as predicted time ago by Luttinger and Tisza,\cite{tisza} and observed recently in
crystals build with NP.\cite{kasutich, dosAF}

{Disordered (non-crystalline) dense }arrays of NP can be obtained from colloidal dispersions of particles in frozen fluids,\cite{ferrofluids} or
by compacting powders of NP in granular solids.\cite{bedanta, powder} The typical volume fractions attained by those systems range from
$20$ to $40\%$. As pointed out by M\o rup,\cite{morup} frozen {\it disorder} in the position of each NP and/or in its random orientation
{together with {\it frustration} (that comes from dipolar interactions) may} result in spin-glass (SG) behavior. \cite{bookstein, bookmezard}
These systems are often called {\it super}-SGs, because they are made of NP. They exhibit typical behavior of SG\cite{luo} such as
anomalous relaxation, aging, and other memory effects similar to the ones previously found in their single-molecule SG counterparts.\cite{binderREV} 
Numerical simulations of SID with different combinations of positional and easy-axis orientational disorder have revealed a
similar behavior,\cite{mcaging, ulrich, labarta, memory, russier, woinska} and also that, irrespective of the relative dipolar interaction
strength, dipoles flip up or down along their local easy axes in an Ising-like manner.\cite{russ}
SG behavior clearly governed by dipolar interactions has been observed
in random closed packed (RCP) samples of highly monodisperse maghemite ($\gamma Fe_{2}O_{3}$) NP.\cite{toro1,toro2} 

Recent Monte Carlo (MC) simulations of diluted systems of either parallel\cite{tam, PADdilu, PADdilu2}  
or randomly oriented Ising dipoles\cite{RADdilu} exhibited the existence of a SG 
phase at temperatures below a transition temperature $T_{sg}$. Moreover, MC data are consistent with quasi-long-range\cite{xy} 
order in the SG phase.\cite{PADdilu, PADdilu2} Neither the droplet model\cite{droplet} nor a replica symmetry breaking (RSB) scenario\cite{RSB} fit in with this marginal behavior. 
Previous MC work for a fully occupied simple cubic (SC) lattice of dipoles with randomly oriented axes\cite{RADjulio} also found a SG phase, but neither
clear-cut results about the nature of the SG phase were obtained, nor was it possible to discern between the above mentioned scenarios.
The validity of {one scenario over the others may depend crucially} on the interactions involved in the system under study.
A paradigmatic model  that exhibits RSB is the Sherrington-Kirkpatrick (SK) model,\cite{sk,solutionSK} 
where the couplings between pairs of spins are chosen randomly to be ferromagnetic or antiferromagnetic regardless of the spin-spin distance.
However, the  applicability of a RSB scenario to more realistic {(short-ranged) }models as the Edwards-Anderson\cite{ea, oldEA} is still  controversial.\cite{bookstein}  

The main purpose of the present work is to investigate by tempered MC simulations the equilibrium SG phase of
dense packings of randomly oriented Ising dipoles.  We consider arrays of dipoles  placed 
on the sites of a fully-occupied SC lattice, and ensembles of dipoles placed at the center of RCP spheres 
that occupy a 64\% fraction of the entire volume. Both packings are  clearly more homogeneous than 
loose-packed configurations with lower volume fraction, or diluted fluid-like positional
configurations in which the number of neighbors changes greatly form particle to particle.

We measure the overlap parameter $q$ between equilibrium configurations, its associated correlation length,
as well as  sample-to-sample fluctuations of probability distributions of $q$ for different realizations
of disorder. 

The paper is organized  as follows. In Sec. \ref{mm} we define the model and the types of dense packings we use,  give details on the MC
algorithm,  and  define the quantities we compute. The results are presented in  Sec.~\ref{results} and some concluding remarks in Sec.~\ref{conclusion}.

\section{models, method, and measured quantities}
\label{mm}
\subsection{Models}
\label{models}
We study the low temperature behavior of dense packings of identical magnetic NP that behave as single magnetic dipoles. 
Each NP $i$ is a hard sphere of diameter $d$
carrying a permanent pointlike magnetic moment  $ \vec{\mu_{i}}=\mu \sigma_{i} \widehat{ a_{i} }$ at its center.
$\widehat{a}_i$ is the local easy-axis and $\sigma_i=\pm1$ is a sign representing the moment $\vec{\mu}_i$
pointing up or down along $\widehat{a}_i$. $\mu\equiv\Vert\vec{\mu}_i\Vert$ is equal for all dipoles.
As in real dense packings of NP, we  consider that $\widehat{a_{i}}$ axes are frozen and point
along randomly distributed directions. Magnetic moments are coupled solely by dipolar interactions. 
The Hamiltonian is given by

\begin{equation}
  {\cal H}=\varepsilon_d \sum_{ <i,j>}  \left( \frac {d}{r_{ij}} \right) ^{3} \left( \widehat{a}_i \cdot \widehat{a}_j-
  \frac {3(\widehat{a}_i\cdot \vec{r}_{ij})( \widehat{a}_j\cdot \vec{r}_{ij})} {r_{ij}^2} \right) \sigma_i \sigma_j
\end{equation}
where $\varepsilon_d =\mu_{0}\mu^2/(4 \pi d^{3})$ is an energy ($\mu_0$ is the magnetic permeability in vacuum), and
the summation runs over all pairs of particles $i$ and $j$ except $i=j$. $\cal H$ can be recast in the manifestly Ising-like form
\begin{equation}
{\cal H}= \sum_{ <i,j>}  T_{ij}  \sigma_i \sigma_j
\end{equation}
where, 
\begin{equation}
  T_{ij}=\varepsilon_d\left( \frac {d}{r_{ij}} \right) ^{3} \left( \widehat{a}_i \cdot \widehat{a}_j-
  \frac {3(\widehat{a}_i\cdot \vec{r}_{ij})( \widehat{a}_j\cdot \vec{r}_{ij})} {r_{ij}^2} \right)\;.
\end{equation}
Since dipoles point along randomly chosen directions $\widehat{a}_i$, $T_{ij}$ signs are distributed at random. Moreover, $T_{ij}$ values depend
on the orientation and on the modulus of the relative position vectors $\vec{r}_{ij}$. 

In our simulations we flip dipoles up and down along their easy axes. Given that we are not interested on time dependent properties controlled
by the interplay between local anisotropy and interparticle dipolar energies, we do not try to mimic how each dipole overcomes anisotropy barriers.
Rather, we try to reproduce the collective evolution effects that follow when the system
is allowed to explore the rough free-energy landscape inherent to SG and relax to equilibrium.

We shall analyze two different types of packings of identical spheres. On the one hand, the spheres have been placed on the nodes of SC lattices
with lattice spacing $d$, and on the other hand, they have been placed in random close packings (RCP). Both
types of packings are collectively called random axial dipole systems (RAD). We do not expect to see relevant differences
in the behavior of the two packings, because both are rather homogeneous and dominated by a random axis distribution.

As for RCP, many experimental results and numerical simulations indicate that in the most compact way, RCP spheres occupy a volume fraction
$\phi=\phi_{0}\equiv0.64$.\cite{torquato}  For that reason, we choose RCP systems with this precise value of $\phi$. 
Note that there is an additional source of randomness in RCP stemming from the spatial disorder in $\vec {r}_{ij}$.

For comparison, we study also the SK model: a set of $N$ Ising spins $\sigma_i=\pm 1$  where any pair of spins interacts. The interaction
energies between the spins at sites $i$ and $j$ is $J_{ij}\sigma_i\sigma_j$ with  $J_{ij}=\pm 1/\sqrt{N}$. The signs in $J_{ij}$ are chosen randomly.

In the following, all temperatures will be given in units of $\varepsilon_{d}/k_{B}$ ($1/k_B$ for the SK model), where $k_{B}$ is Botzmann's constant.

\subsection{Method}

We have simulated $N_{s}$ independent {\it samples} of the above-described models.
A {\it sample} $\cal J$ is a given realization of quenched disorder. For RAD systems this disorder means
choosing  the orientations of vectors $\widehat{a}_i$ randomly, while a sample for the SK model is defined by the distribution of signs in $J_{ij}$.
Besides the disorder in the orientations of $\widehat{a}_i$, RCP systems include another source of disorder, namely the positions $\vec{r}_i$ of the spheres.
This second cause of disorder is absent in SC systems. 

To fix $\vec{r}_i$ in RCP systems, the Lubachevsky-Stillinger (LS) algorithm\cite{ls, donev} has been used. With it, a system of $N$ identical hard spheres
evolve according to Newtonian dynamics. At the same time, all the particles are let to grow in size. Furthermore, this growth is performed  at a sufficiently
high rate in order to avoid the system ending up in a crystaline state. Proceeding in this way, at the end the system gets up stuck in a disordered
state.\cite{torquato, donev} To be precise, we start placing a Poisson distribution of $N$ small spheres by random sequential addition in a cube whose
edges have length $L=1$. At the beginning the spheres occupy a volume fraction $\phi=0.2$. Periodic boundary condition are applied. The LS algorithm
lets spheres move freely and grow until the sample eventually reaches the volume fraction $\phi_{0}$.\cite{donev} Finally, once the LS algorithm has
stopped evolving the spheres and their size, positions $\vec r_{i}$ are rescaled
such that all spheres recover a diameter $d=1$ and $L$ becomes $L= (N \pi/6\phi_{0})^{1/3}$.  

The system size is determined by the number $N$ of spheres. The number of samples $N_{s}$ is listed in Tables~I and~II for every size $N$.
In SG systems  statistical errors are independent  of $N$ because of their inherent  lack of self-averaging. This is why we have not made $N_s$
smaller with increasing $N$. For RAD systems in SC lattices with $N=1000$, we could only employ $1400$ samples because of CPU time limitations.

Periodic boundary conditions are always used. We let each dipole $i$ interact with the other dipoles within an $L\times L\times L$
cube centered on $i$ and with the repeated copies of the dipoles beyond the box (by periodicity).
In order to take into account the slowly decaying long-range dipole-dipole interaction we do perform Ewalds's summations.\cite{ewald, allen} We follow 
the notation from the paper by Wang and Holm.\cite{holm} In this method, pointlike dipoles are screened by Gaussians with standard deviation $1/2\alpha$ 
that allow to split the computation of the dipolar energy into two rapidly convergent sums, one
in real space and the other in reciprocal space. We evaluate the sum in real space
using the normal image convention, with a cutoff $r_{c}=L/2$. Also a reciprocal space cutoff $k_{c}$ is introduced for the sum in the reciprocal space.  
We have chosen $k_{c}=10$,  and  $\alpha=7.9/L$ as a good compromise between accuracy and calculation speed.\cite{holm}   Finally, given that our
system is  expected to exhibit zero magnetization, we have used a surrounding permeability $\mu'=1$.

In order to  reach equilibrium at low temperatures in the SG phase 
we use  a parallel tempered Monte Carlo (TMC) algorithm.\cite{tempered}
It consists in running a set of $n$ identical replicas of each sample in parallel at different 
temperatures in the interval $[T_{min},T_{max}]$ with a separation $\Delta$ between neighboring temperatures.
Each replica starts from a completely disordered configuration $\{\sigma_{i}\}$.  
We apply the TMC algorithm in two steps. In the first one, the $n$ replicas of the
sample $\cal J$ evolve  independently for $8$ Metropolis sweeps.\cite{mc}  All dipolar fields throughout the system are
updated every time a spin flip is accepted. In the second step, 
we give to any pair of replicas evolving at temperatures $T$ and $T-\Delta$ a chance to exchange states
between them following standard tempering rules which  satisfy detailed balance.\cite{tempered}
These exchanges allow all replicas to diffuse back and forth from low to high temperatures 
and reduce equilibration times for the rough energy landscapes expected for SGs.
We find it helpful to choose $T_{max}$ larger than $2 \times T_{sg}$ and choose $\Delta$ such that at least $30 \%$ of all 
attempted exchanges are accepted for all pairs $(T, T-\Delta)$.

Measurements were taken after two averagings: firstly over thermalized states of a given sample $\cal J$ and secondly over  $N_{s}$ samples with different
realizations of quenched disorder.  Thermal averages come from averaging over the time interval $[t_0,2 t_0]$, where $t_{0}$ is the equilibration
time. Given an observable $\verb|u|$, $u_{\cal J}=\langle~ \verb|u|~  \rangle_{T}$ stands for the thermal 
average of sample $\cal J$ and $\langle u\rangle= [~u_{\cal J~}]_{\cal J}$ for the average over samples.
The values of various parameters used in the simulation runs are given in Tables~I and~II.

\begin{table}[!h]
\begin{tabular}{p{1.2cm} p{1cm } p{1cm } p{1cm } p{1.4cm } p{1.4cm }}
\hline
\hline
\multicolumn{6}{c}
{Simple Cubic}\\
\hline
$N$ & $T_{min}$ & $T_{max}$ & $\Delta$ & $t_{0}$ &$N_{s}$\\
\hline
$64$ &  $0.2$        & $2.1$         & $0.05$ &  $8 \times10^{6}$ & $5100$ \\
$125$&  $0.2$       & $2.1$	& $0.05$ &   $8 \times10^{6}$ & $10000$ \\
$216$&  $0.2$       & $2.1$	& $0.05$ &   $8 \times10^{6}$ & $10000$ \\
$343$&  $0.2$       & $2.1$	& $0.05$ &   $8 \times10^{6}$ & $4800$ \\
$512$&  $0.2$       & $2.1$	& $0.05$ &   $8 \times10^{6}$ & $5100$ \\
$1000$& $0.6$      & $2.1$	& $0.05$ &   $10^{7}$ & $1400$\\
\hline
\multicolumn{6}{c}
{Random Close Packaged}\\
\hline
$N$ & $T_{min}$ & $T_{max}$ & $\Delta$ & $t_{0}$ &$N_{s}$\\
\hline
$125$ &  $0.2$   & $2.1$ &   $0.05$ &$8\times10^{6}$ & $5900$ \\
$216$ &  $0.2$   & $2.1$ &   $0.05$ &$8\times10^{6}$ & $8000$ \\
$512$ &  $0.2$   & $2.1$ &   $0.05$ &$8\times10^{6}$ & $7700$ \\
\hline
\hline
\end{tabular}
\caption{ Simulation parameters for SC and RCP systems. $N$ is the number of dipoles, $T_{min}$ 
($T_{max}$) is the lowest (highest) temperature and $\Delta$ is the temperature step in our TMC
simulations. The number of simulation sweeps for equilibration is $t_{0}$. Measurements are taken 
in the time interval $[t_{0}, 2t_{0}]$. The number of samples with different realizations of 
quenched disorder is $N_{s}$.} 
\label{table1}
\end{table}

\begin{table}[!h]
\begin{tabular}{p{0.8cm} p{1cm } p{1cm } p{1cm } p{1.4cm } p{1.4cm }}
\hline
\hline
$N$ & $T_{min}$ & $T_{max}$ & $\Delta$ & $t_{0}$ &$N_{s}$\\
\hline
$64$ &  $0.16$       & $1.60$         & $0.04$ & $10^{5}$ & $10^{5}$ \\
$216$ &  $0.16$       & $1.60$         & $0.04$ & $10^{5}$ & $1.4\times 10^{5}$ \\
$512$ &  $0.16$       & $1.60$         & $0.04$ & $2\times10^{5}$ & $10^{5}$ \\
\hline
\hline
\end{tabular}
\caption{Same as in Table I for the SK model.}
\label{tablaSK}
\end{table}

\subsection{Observables}
\label{meas}

The SG behavior has been investigated by measuring the spin overlap parameter,\cite{ea}
\begin{equation} 
q\equiv N^{-1} \sum_j \sigma^{(1)}_j\sigma^{(2)}_j ,
\label{q}
\end{equation}
where $\sigma^{(1)}_j$ and $\sigma^{(2)}_j$ are the spins on site $j$ of two independent equilibrium states, called
$(1)$ and $(2)$, of a given sample. Clearly,  $q$  is a measure of the spin configuration overlap between 
the two states. To avoid unwanted correlations, we do not look for states (1) and (2) in single samples.
Rather, we consider for each sample, a pair of identical replicas that evolve independently in time.

We evaluate the order parameters $q_{2}\equiv \langle~ q^{2}~ \rangle$ and
$q_{1}\equiv \langle~ |q|~  \rangle $ and, for each sample $\cal J$ the overlap probability distribution $p_{\cal J}(q)$. Then, the
mean overlap distribution $p(q)$ over all replicas is defined as
\begin{equation} 
p(q) \equiv  [~ p_{\cal J}(q) ~]_{\cal J}.
\label{pq}
\end{equation}
We also measure the mean square deviations of $p_{\cal J}(q)$, from the average $p(q)$,
\begin{equation} 
\delta p(q)^{2} \equiv  [~ \{p_{\cal J}(q) -p(q)\}^{2}~ ]_{\cal J} .
\label{dq}
\end{equation}

As usual in SG work,\cite{longi,balle,katz0} the correlation length $\xi_L$ is computed by
\begin{equation} 
\xi^2_L\equiv\frac {1 } {4 \sin^2  ( k /2)}  { \left \{ \frac{\langle q^2 \rangle} { \langle\mid q({\bf k})  \mid ^2   \rangle}  -1 \right \} }, 
\label{phi1}
\end{equation}
where
\begin{equation} 
q({\bf k})\equiv N^{-1} \sum_j \psi_j e^{{\rm i} \vec{k}\cdot \vec{r}_j},
\label{phi2}
\end{equation}
with $\psi_j\equiv\sigma^{(1)}_j\sigma^{(2)}_j$, $\vec{r}_j$ the position of dipole $j$, and $\vec{k} =(2\pi/L,0,0)$
and $k=\Vert\vec{k}\Vert=2\pi/L$. Recall that since our systems are isotropic, all directions $\vec{k}$ are equivalent.

The correlation function $\langle\psi_r\psi_0\rangle-\langle\psi_r\rangle\langle\psi_0\rangle$ decays as $\exp(-r/\xi_\infty)$ where
$\xi_\infty$ is the correlation length in the thermodynamic limit. $\xi_L$ in (\ref{phi1}) provides a good approximation of $\xi_\infty$ in the
$\xi_L/L\to0$ limit in the paramagnetic phase for which $\langle\psi_r\rangle$ vanishes.\cite{balle}

This is not the case  for an ordered phase. Consider for example strong long--range  order with
short--range order fluctuations. That is,  $\langle \psi_0\psi_r\rangle$ does not vanish as $r\to \infty$ and
only $\langle \psi_0\psi_r \rangle -\langle \psi_0\rangle  \langle\psi_r \rangle$ is short--range.
In such a case, the ratio $(\xi_L/L)^{2}$ diverges as $L^3$ as $L$ increases, and is not related
with ${\xi_\infty}$.\cite{balle}
One would have to replace $\psi$ by $\psi -\langle \psi\rangle$  in Eq.~(\ref{phi1}) in order to relate $\xi_L$ to $\xi_\infty$ in the thermodynamic limit.
Following current usage, we shall nevertheless refer to $\xi_L$ as ``the correlation length''.

Note that, in contrast with $p(q)$ and its first moments, $\xi_{L}$ takes into account spatial variations of the
overlap $q$.

\subsection{Equilibration Times}

It is important to make sure that thermal equilibrium is reached before we start taking measures. To do that, we followed the same procedure as in Ref.~\onlinecite{PADdilu2},
by defining, for two replicas of a single sample, the overlap $q_{t}$  at time $t$ and the average $q_{2}(t)$ of its square over all samples. Equilibrium is
reached when $q_{2}(t)$ attains a plateau. In order to confirm this result, a second overlap $\widetilde q_{t_{0},t}$
\begin{equation} 
\tilde q_{t_0,t}\equiv N^{-1} \sum_j \sigma_j(t_0)\sigma_j( t_0+t)\;,
\label{phi0}
\end{equation}
between spin configurations of a single replica taken at times $t_0$ and $t_0+t$ is measured as a function of $t$.
Equilibrium imposes that the corresponding average $\widetilde q_2(t_{0},t)$ remains stuck to the above plateau as $t$ varies.\cite{PADdilu2}


\begin{figure}[!t]
\includegraphics*[width=78mm]{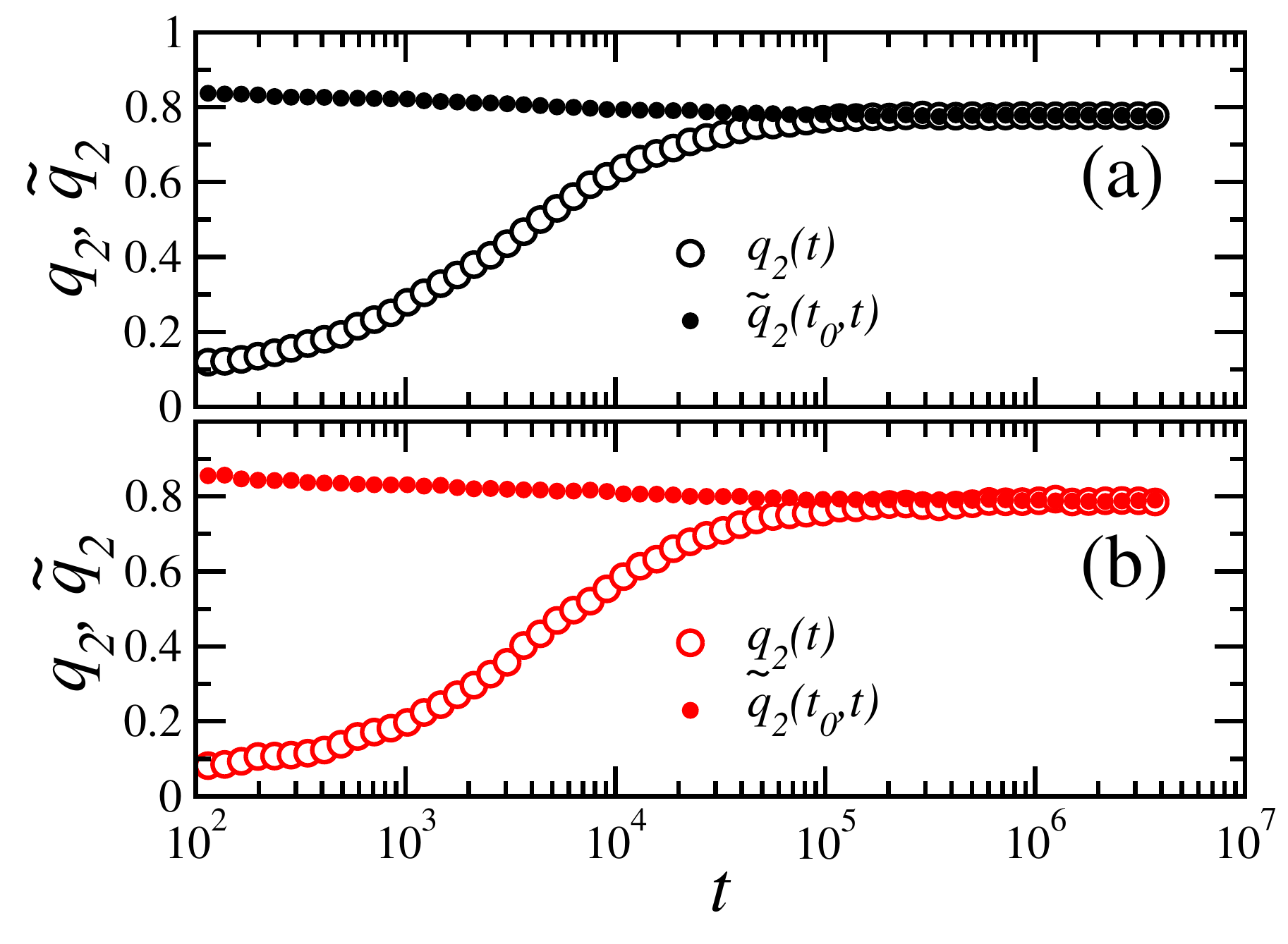}
\caption{(Color online)
  (a) Semilog plots of $\widetilde q_2(t_0,t)$ ($\circ$) and $q_2(t)$ ($\bullet$) vs. time $t$ (measured in Metropolis sweeps) for SC
  packings of $512$ dipoles evolving at $T=0.2$, the lowest temperature of our TMC simulations.  
  Here, $t_0=8 \times10^6$ Metropolis sweeps. Data points at time $t$ stand for an average over the time interval $[t,1.2t]$, and over $10^{3}$ samples. (b) The same
  for RCP.
}
\label{tiempos}
\end{figure}


Plots of $\tilde{q}_2(t_0,t)$ and $q_2(t)$ vs. $t$ are shown in Fig.~\ref{tiempos}(a) (Fig.~\ref{tiempos}(b)) for
RAD systems on SC lattices (RCP) for $t_0=5\times 10^6$ Metropolis sweeps with $N=512$ and $T=0.2$, the lowest value of $T$ in the series of
TMC simulations. 
We have chosen sufficiently large values  of $t_0$ to make sure that $\tilde{q}_2(t_0,t)\approx q_2(t)$ for  $t\gtrsim t_{0}$.

After letting the system equilibrate for a time $t_{0}$, we
have taken averages over the time interval $[t_{0},2t_{0}]$. The values of $t_0$ and $N_s$ employed in our runs are given in Tables~I and~II.

It has  been shown that equilibration times of individual samples
are directly correlated with the roughness of their free-energy landscape.\cite{yuce2}  
Numerous spikes in the overlap distributions $p_{\cal J}$ are associated to samples that have
several pure states.\cite{aspelmeier} The symmetry of the plots of overlap distributions like the ones 
shown in Fig.~\ref{p_delta}(a) for SC are an additional check that all the samples are
well equilibrated.

\section{RESULTS}
\label{results}

\subsection{The SG Phase}
\label{SGphase}

In this section, we report numerical results for $q_2$ and $\xi_L/L$.

\begin{figure}[!b]
\includegraphics*[width=75mm]{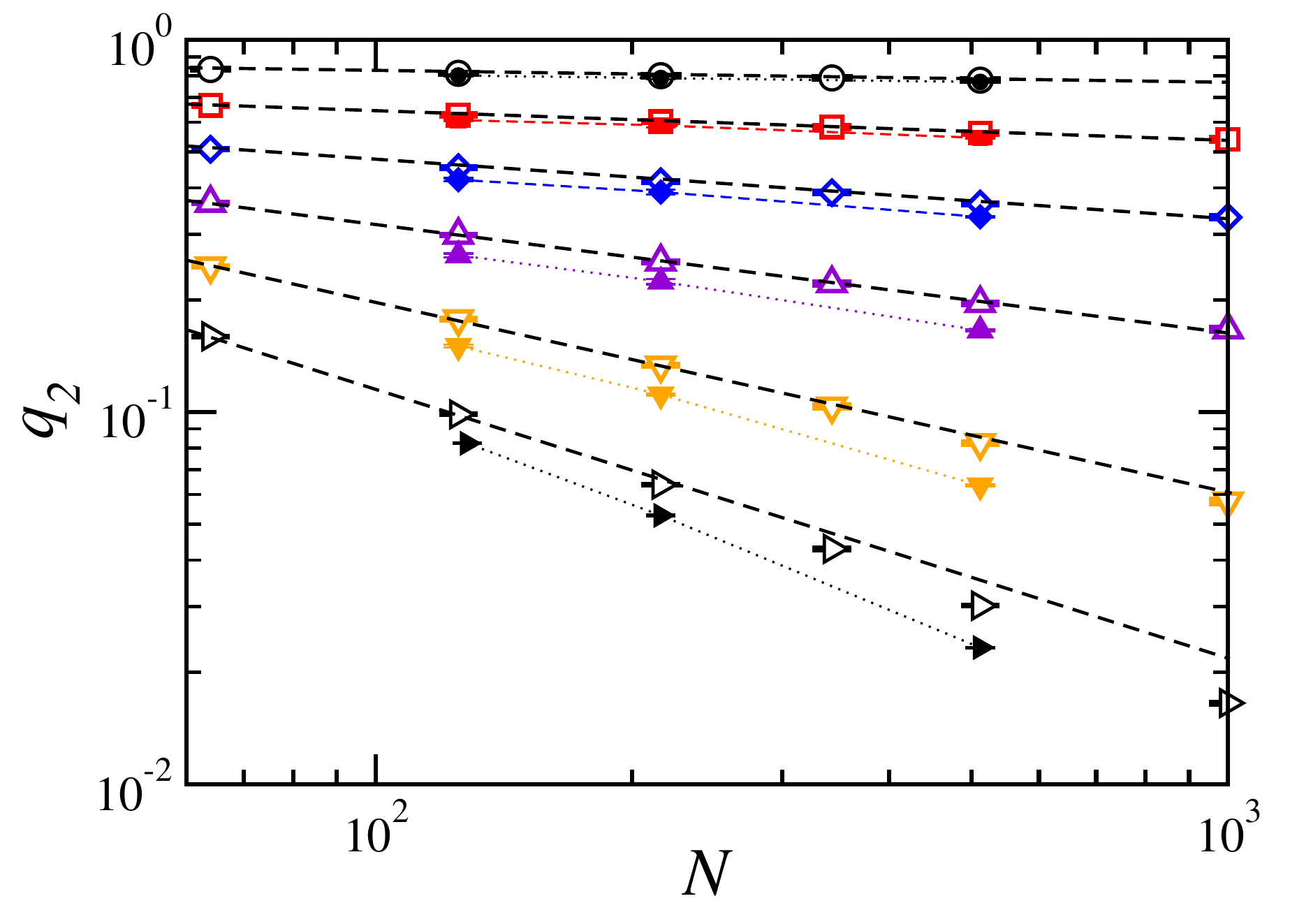}
\caption{ 
 (Color online) Log-log plots of $q_2$ vs. the number of dipoles $N$ for SC and RCP.
$\circ$, $\square$, $\diamond$, $\triangle$, $\triangledown$, and $\triangleright$
stand for SC systems at temperatures $T=0.2, 0.4, 0.6, 0.8, 1.0$, and $1.2$ respectively.
$\bullet$, $\blacksquare$, $\blacklozenge$, $\blacktriangle$, $\blacktriangledown$, and $\blacktriangleright$
stand for RCP systems at the same temperatures.
Dotted lines are guides to the eye. Data sets for larger temperatures deviate from a linear trend (represented by the straight dashed lines),
implying a decay faster than a power of $1/N$. Data for the lower lower temperatures are well fitted by the straight lines.
}
\label{q_vs_L}
\end{figure}

A log-log plot of  $q_2$ vs. $N$ for different values of $T$ is shown in Fig.~\ref{q_vs_L} for RAD systems in SC
lattices and RCP arrangements. For both models $q_2$ decreases as $N$ increases, and this occurs at all temperatures.  
For $T<0.8$ and the system sizes studied, data points in this figure are consistent with
an algebraic decay $q_2\sim N^{-(1+\eta)}$, following the usual definition of exponent $\eta$.\cite{mefisher} 
Plots of  $q_1$  vs. $N$ (not shown) exhibit the same qualitative behavior.
All of this is in accordance with quasi-long-range order. Previous simulations\cite{RADjulio} for the SC were not able to discriminate between the
scenario where $q_2$ tends to a constant value as $N$ grows and the algebraic steady decay shown in Fig.~\ref{q_vs_L}.
We will return to this point below. From the plots for SC systems, one could extract $\eta$ for various values of $T$. The relation 
$\eta =-1+a~T^2$ fits the data well with $a=0.45$. Our results disagree with  a RSB scenario,\cite{RSB} in which $q_2$ does not vanish as $L \to \infty$.\cite{remark} 
For even higher temperatures than those shown in Fig.~\ref{q_vs_L},  $q_2$ vs. $N$  curves bend downwards, as expected for
the paramagnetic phase. Approximate values of $T_{sg}$ can thus be obtained from such plots, but more accurate methods are given below. 

\begin{figure}[!t]
\includegraphics*[width=88mm]{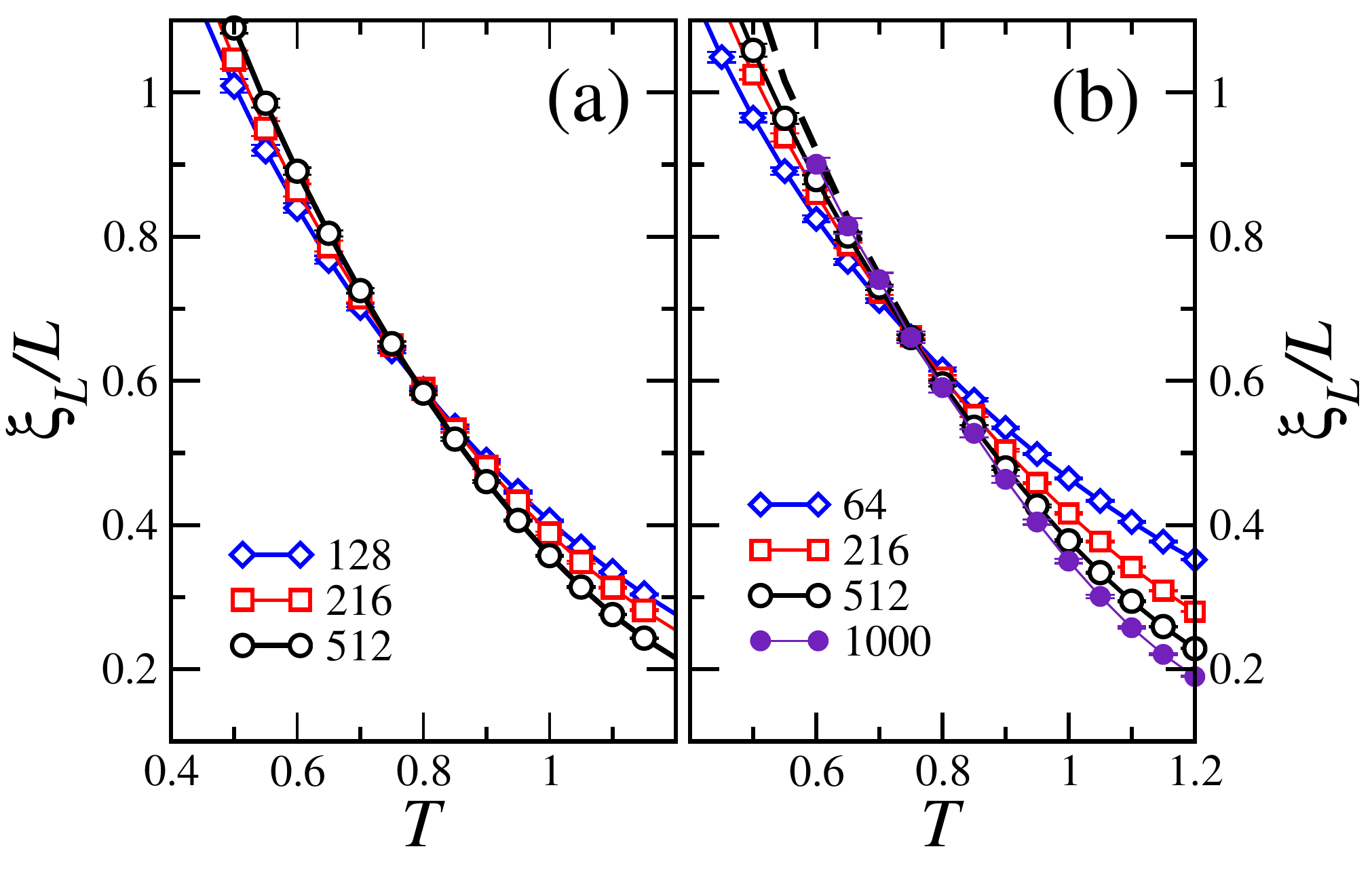}
\caption{(Color online) 
(a) Plots of $\xi_L /L$ vs. $T$ for RCP systems with the number of dipoles $N$ indicated in the figure. 
Recall that dipoles are placed on a cube of volume $L^{3}=N\pi/6\phi$, where $\phi$ is the volume fraction
occupied by the spheres. Continuous lines are guides to the eye. 
(b) Same as in (a) but for SC systems. Now,  $L=N^{1/3}$.
The thick dashed line follows from the $1/L\to 0$ linear extrapolations in the plots of Fig. \ref{longs_vs_L} for $T<T_{sg}$.
}
\label{longsBIS}
\end{figure}


We next examine how  $\xi_L/L$ behaves for RAD systems in dense packings with SC and RCP. This
has already been done for diluted systems of parallel Ising dipoles.\cite{tam, PADdilu}
We aim at exploring the behavior  of $\xi_L/L$ not only near 
$T_{sg}$, but also deep into the SG phase.  Recall that $\xi_L$ becomes a true correlation 
length in the paramagnetic phase when $\xi _L/L\ll 1$. Then, $\xi_L/L$ falls off as $1/L$ in this phase.
On the contrary,  $\xi _L/L$ increases with  $L$ in the SG phase.
The system passes from one phase to the other at a temperature
$T_{sg}$, so that we can reasonably expect that curves of $\xi _L/L$ vs. $T$ for different values of $N$ cross at $T_{sg}$.
All that enables us to extract
$T_{sg}$ from those plots. At $T=T_{sg}$, $\xi_L/L$ must become size independent, as expected for a scale free quantity.

Plots of $\xi_L/L$ vs.  $T$ are shown in Fig.~\ref{longsBIS} for different values of $N$ on SC (Fig.~\ref{longsBIS}(a)) and RCP (Fig.~\ref{longsBIS}(b))
arrays. All curves spread out above and below  a quite precise crossing point $T_{sg}$ and this fact
allows to obtain a precise determination of $T_{sg}$ from the intersection of $\xi_L/L$ vs. $T$ curves
as is sometimes done for the EA \cite{longi,balle,katz0} and dipolar SG\cite{tam, PADdilu} models.
For our SC (RCP) systems, curves cross at $T_{sg}=0.75(2)$ ($T_{sg}=0.78(3)$). 

We now focus on the data for $\xi_L$ at low temperatures, and check if they are consistent
with the algebraic decay of $\langle~q^2~\rangle$ exhibited in Fig.~\ref{q_vs_L}.

\begin{figure}[!t]
\includegraphics*[width=78mm]{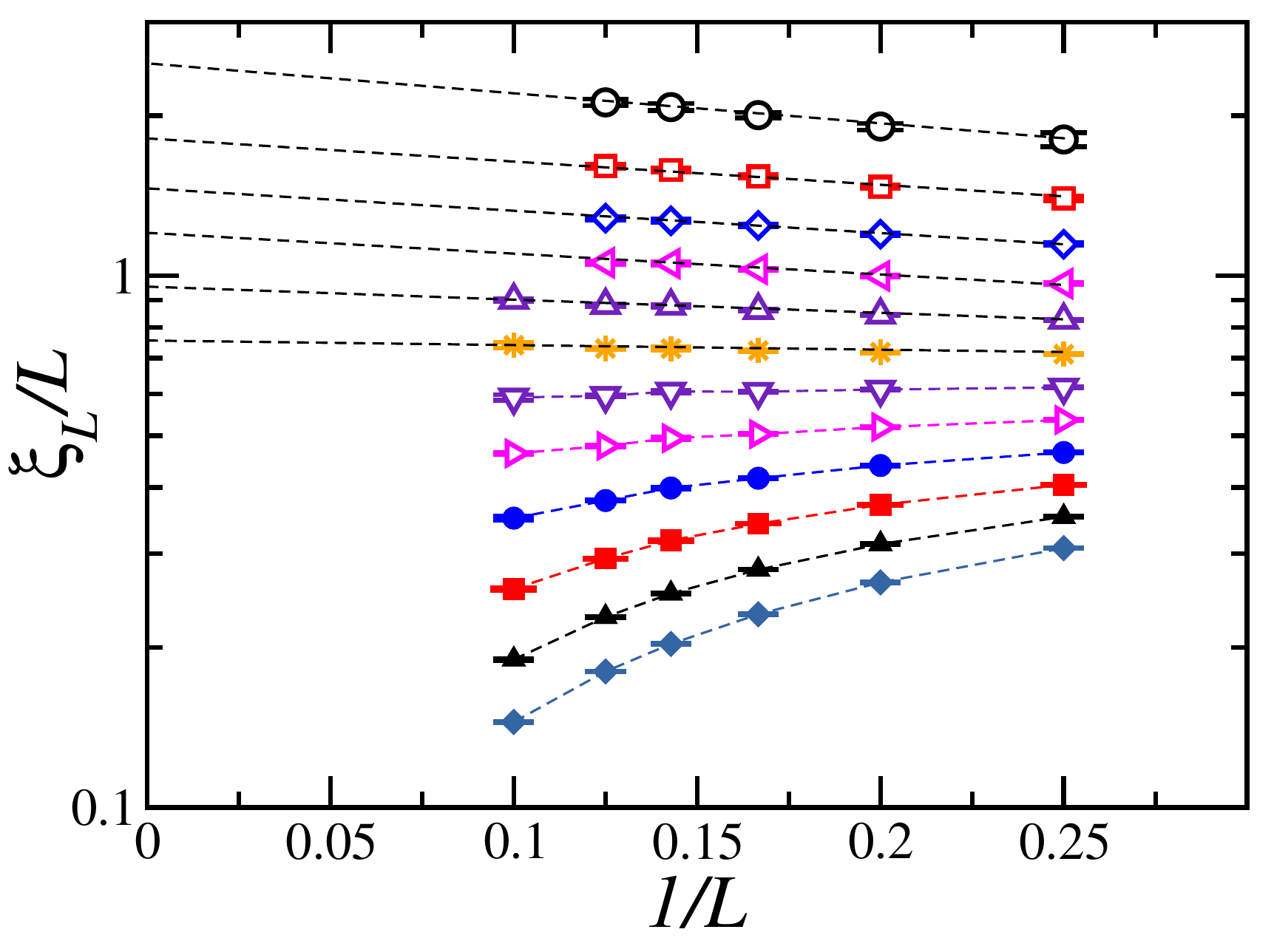}
\caption{ (Color online) 
Semilog plots of $\xi_L/L$ vs. $1/L$ for SC systems, and  
$T=0.2$ ($\circ$), $T=0.3$ ($\square$), $T=0.4$ ($\diamond$), $T=0.5$ ($\triangleleft$), $T=0.6$ ($\triangle$),
 $T=0.7$ ($\ast$), $T=0.8$ ($\triangledown$), $T=0.9$ ($\triangleright$), $T=1.0$ ($\bullet$), $T=1.1$ ($\blacksquare$),$T=1.2$ ($\blacktriangle$), and $T=1.3$ ($\blacklozenge$). 
Dashed lines are guides to the eye. 
}
\label{longs_vs_L}
\end{figure}

In the droplet model picture, in the SG phase $q_2\neq 0$ and $\langle \psi_0\psi_r \rangle -\langle \psi_0\rangle  \langle\psi_r \rangle$ is short ranged.
\cite{droplet} It then follows that\cite{PADdilu} $\xi_L^2/L^2\sim L^3$.
No feature exists in the plots of $\xi_L/L$ vs. $1/L$ shown in Fig. ~\ref{longs_vs_L} suggesting that trend, and this is valid at all temperatures.

Let us assume now that that the connected correlation function 
$\langle \psi_0\psi_r \rangle -\langle \psi_0\rangle  \langle\psi_r \rangle$ decays as $G(r) \sim 1/r^{(1+\eta)}$
as $r \to \infty$ while having $q_2\not=0$. This behavior fits in with the RSB picture.\cite{RSB}
Then, it follows from Eq.~(\ref {phi1}) that  $\xi_L^2/L^2\sim L^{1+\eta}$.\cite{PADdilu} 
Neither evidence for { $\xi_L^2/L^2\sim L^{1+\eta}$ } does appear in Fig.~\ref{longs_vs_L}.
Note that the values of $\xi_L/L$ diminish as a function of $1/L$ for  $T < T_{sg}$ and the downward trend becomes steeper as $T$ decreases.
On the other hand, from Fig.~\ref{q_vs_L} we know that $\mid 1+\eta \mid$ decreases with $T$.
This would lead to $\xi_L/L$ vs. $1/L$ curves which do not become steeper as $T$ decreases, which is in clear contradiction with plots in Fig. \ref{longs_vs_L}. 

Finally, let us consider $q_2 = 0$ and $\langle \psi_0\psi_r \rangle =G(r)$ as in the 2D $XY$ model.\cite{xy}
It then follows that $\xi_L/L$ becomes independent of $L$ for large $L$. This is  the outcome from $1/L\to 0$
extrapolations of the dashed straight lines shown in Fig.~\ref{longs_vs_L} for $T\lesssim T_{sg}$.  Thus, a straightforward
interpretation of the data shown in Fig.~\ref{longs_vs_L} is that the SG phase for our densely packed RAD systems behaves marginally.\cite{remark2}


\subsection{Overlap Distributions}
\label{OD}

\begin{figure}[!t]
\includegraphics*[width=70mm]{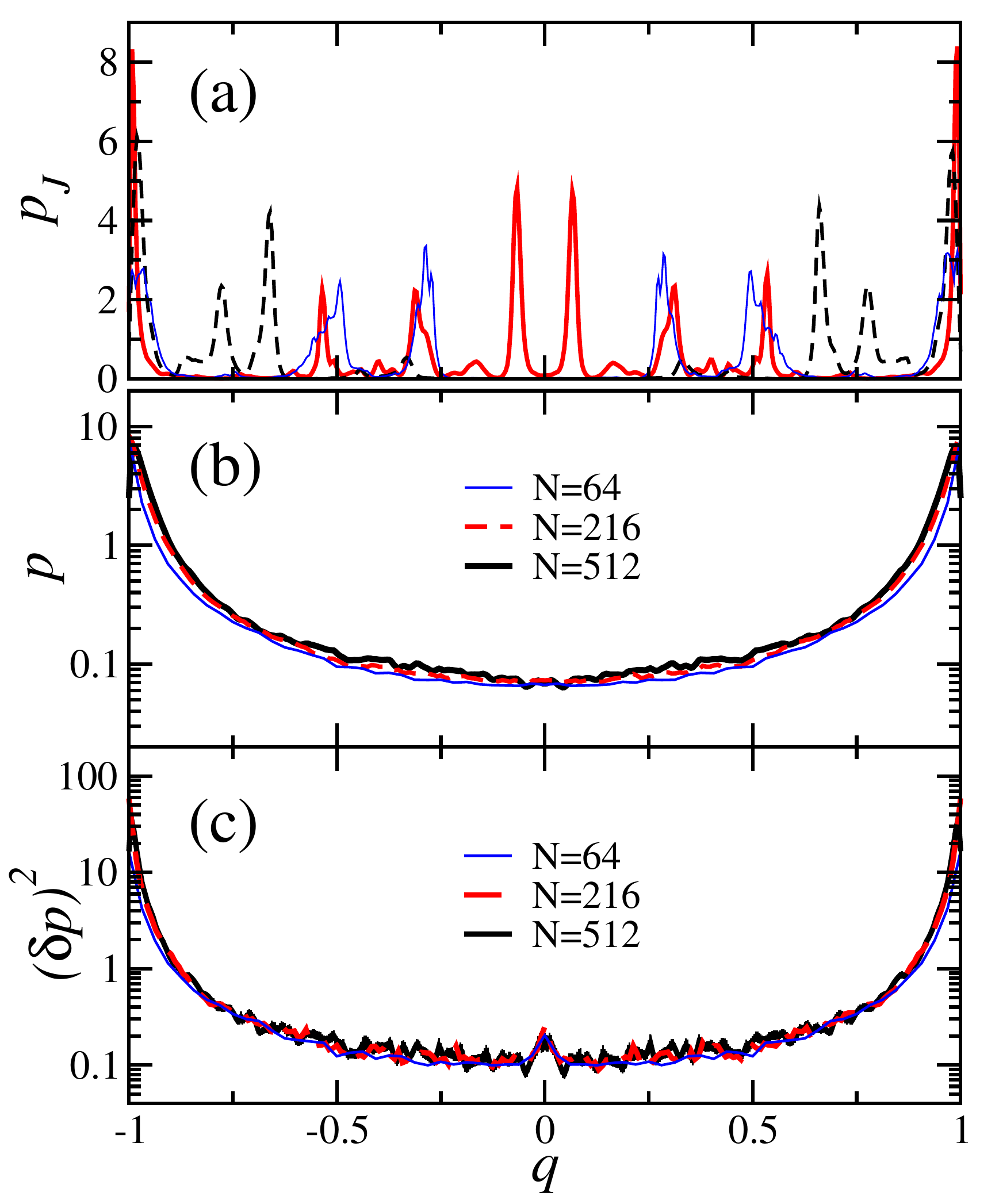}
\caption{(Color online) (a) Overlap distributions $p_{\cal J}(q)$ for SC systems with $N=512$  and $T=0.2$ for three samples with different realizations
of disorder (b) Plots of the averaged distribution $p(q)$ vs. $q$ for SC systems with 
$T=0.2$, and the values of $N$ shown. (c) Same as in (b) for $(\delta p)^{2}$. Similar results obtain for RCP systems.}
\label{p_delta}
\end{figure}

It is interesting to study the SG behavior of individual samples. In Fig.~\ref{p_delta}(a) we plot  $p_{\cal J}(q)$ vs. $q$ for three
different samples at temperature $T=0.2$ for RADs on a SC lattice.  Note that $p_{\cal J}(q)$ distributions are markedly sample-dependent
and exhibit several sharp spikes centered well away from $q \approx \pm 1$. Similar behavior has been found for the EA and SK models.\cite{aspelmeier, yucesoy}
The positions and heights of the spikes in the region $q \in (-Q,Q)$ for, say, $Q\approx 1/2$ change greatly from sample to  sample.
These inner peaks arise from cross overlaps between different pure states. We name these spikes \emph{cross-overlap} (CO) spikes. 
Their number in $p_{\cal J}(q)$ is closely related to the number of pure states. \cite{aspelmeier}
At higher temperatures (not shown), thermal fluctuations render individual spikes so wide that they overlap and become
not clearly discernible. Then, in order to examine CO spikes we were compelled to choose $T_{min} \approx T_{sg}/4$ in our simulations. 

Figure \ref{p_delta}(b) shows the mean overlap distribution $p(q)$ for 
SC arrays at  $T=0.2$ for different values of $N$.  $p(q)$ exhibits two large peaks at $\pm q_{m}$ with $q_{m}\approx1$ that correspond to the self overlap of pure states. We find that $p(q)$ is
approximately flat in the region of small values of $q$,  
with $p(0)\ne0$. 
Note that $p(0)$ is essentially independent of $N$, as previously found
for the SK and EA models.\cite{solutionSK, kpy} This behavior is in conflict with the 
droplet picture of SGs, for which $p(0)\sim N^{-\Theta}$.\cite{droplet}

Plots of  $(\delta p)^2$ vs. $q$ are shown for SC arrays at the same temperature 
in Fig. \ref{p_delta}(c). We obtain (not shown) qualitatively similar plots for RCP systems.
$(\delta p)^2$, which is a measure of sample-to-sample fluctuations of $p_{\cal J}(q)$  from the average $p(q)$, 
does not change appreciably with $N$. According to the 
RSB scenario, $p_{\cal J}(q)$ should exhibit many CO spikes 
that become Dirac delta functions as $N$ increases, bringing about a diverging $(\delta p)^2$
in the thermodynamic limit. 

In order to improve the accuracy,
we consider the integrated probability functions
\begin{equation} 
X_{\cal J}^{Q} \equiv   \int_{-Q}^{+Q} p_{\cal J}(q) dq,
\label{xq}
\end{equation}
\begin{equation} 
\Delta _{\cal J}^{Q} \equiv  \big( \int_{-Q}^{+Q} \{p_{\cal J}(q) -p(q)\}^{2} dq~\big)^{1/2},
\label{Deltaq}
\end{equation}
and compute their related sample averages $X^{Q}$ and $\Delta^{Q}$.

\begin{figure}[!b]
\includegraphics*[width=89mm]{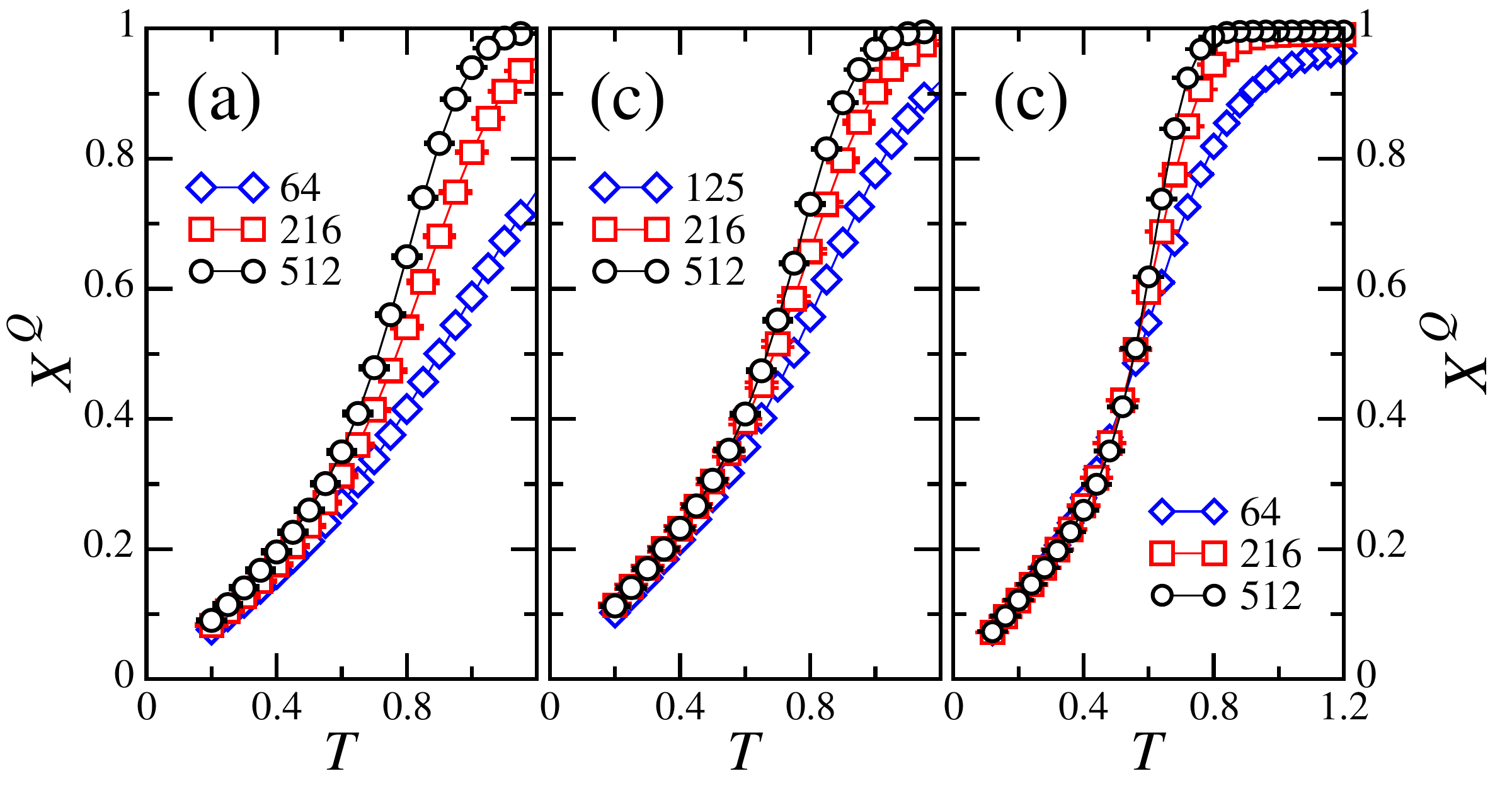}
\caption{(Color online) (a) Plots of ${X}^{Q}$ vs. $T$ for SC systems, $Q=1/2$ and the values of 
$N$ indicated in the figure. (b) The same plots for RCP systems. (c) The same plots for the SK model.}
\label{pvsT}
\end{figure}

Plots of $X^{Q}$ vs. $T$ are shown for $Q=1/2$  in Figs.~\ref{pvsT}(a),  \ref{pvsT}(b), and \ref{pvsT}(c) for SC, RCP systems and the SK model 
respectively. In all cases, $X^{Q}$ appears to be  size independent at temperatures well below $T_{sg}$,
in agreement with mean field predictions for the SK model. 
 This is in clear contrast to the droplet picture, for which $X^{Q}$ is predicted to vanish as $N^{-\Theta}$.\cite{droplet} 
 Finally, we note that in all cases $X^{Q}\propto T$ at low temperatures.

\begin{figure}[!t]
\includegraphics*[width=89mm]{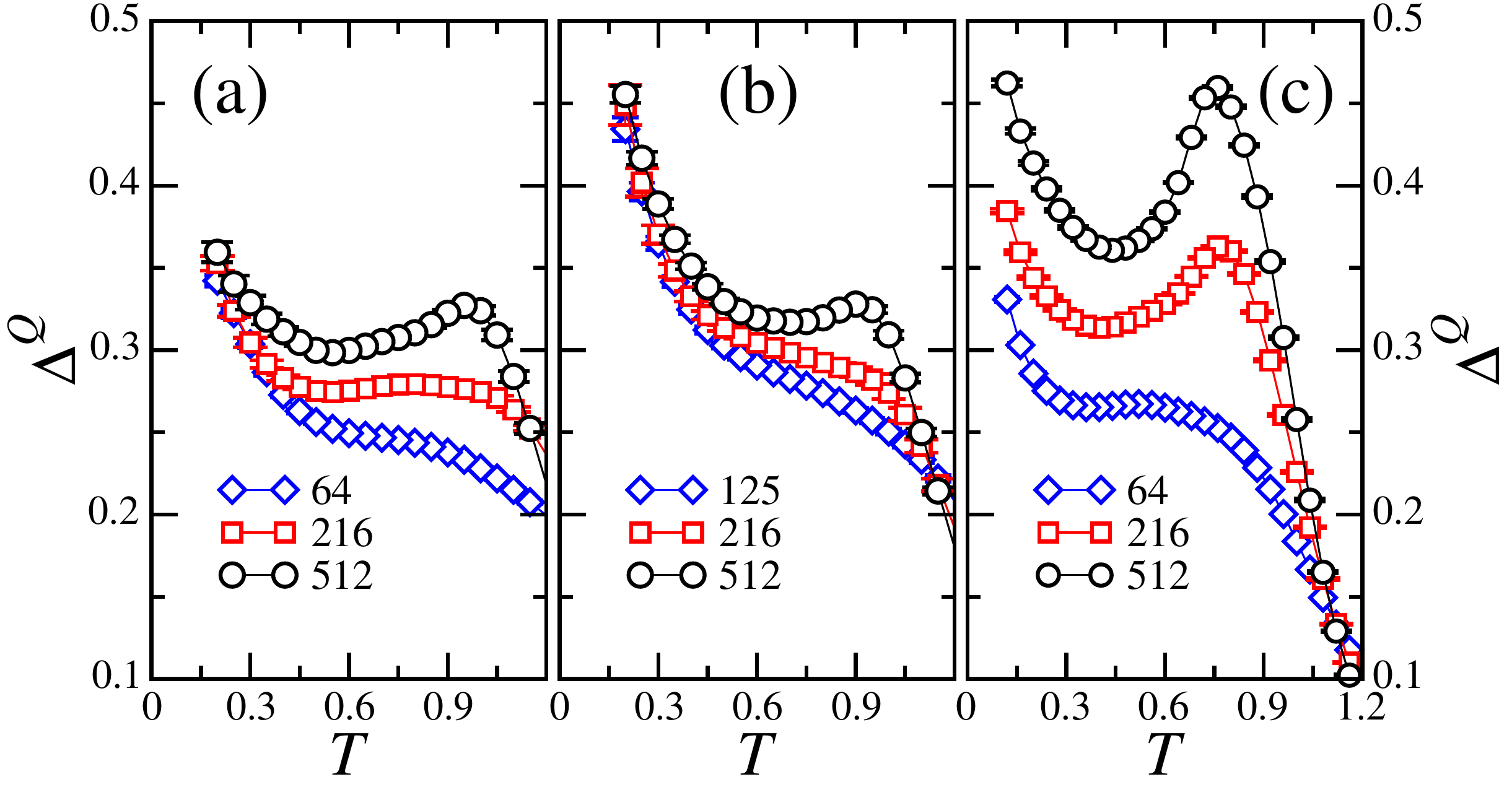}
\caption{(Color online) (a) Plots of $\Delta_{Q}$ vs. $T$ for SC systems, with $Q=1/2$ and the values of 
$N$ indicated in the figure.  (b) The same plots for RCP systems. (c) The same plots for the SK model.}
\label{DeltavsT}
\end{figure}

Plots of $\Delta^{Q}$ vs. $T$ for RADs in Figs.~\ref{DeltavsT}(a) and \ref{DeltavsT}(b) for SC and RCP
suggest that $\Delta^{Q}$ does not diverge as $N$ increases at very low temperatures.
In marked contrast, the corresponding plots for the SK model exhibited  in Fig.~\ref{DeltavsT}(c) 
show that $\Delta^{Q}$ clearly increases with $N$  at all temperatures below $T_{sg}$, in
agreement with the RSB scenario.

\subsection{Cross Overlap Spikes}
\label{pcf}
The shape and width of CO spikes from a \emph{pair correlation function} was studied in previous work.\cite{pcf}
With $f_{\cal J}(q_1,q_2)\equiv   p_{\cal J}(q_1)p_{\cal J}(q_2) $, we define
\begin{equation}
G_{\cal J}^{~Q}(q)  \equiv \int_{0}^{Q} \int_{0}^{Q} dq_1 dq_2\; \delta (q_2-q_1-q)f_{\cal J}(q_1,q_2),
\label{GG}
\end{equation}
and $G^{Q}(q)$ as the average of $G_{\cal J}^{~Q}(q) $ over samples. 
We have computed the normalized function
\begin{equation}
g^{Q}(q)  \equiv\frac{G^{~Q}(q)}{\displaystyle{\int_{-Q}^{+Q} G^{Q}(q) ~dq}}\;,
\label{g}
\end{equation}
which is the conditional probability density that  $q=q_2-q_1$, given that $q_1,q_2\in (0,Q)$. 

\begin{figure}[!t]
\includegraphics*[width=89mm]{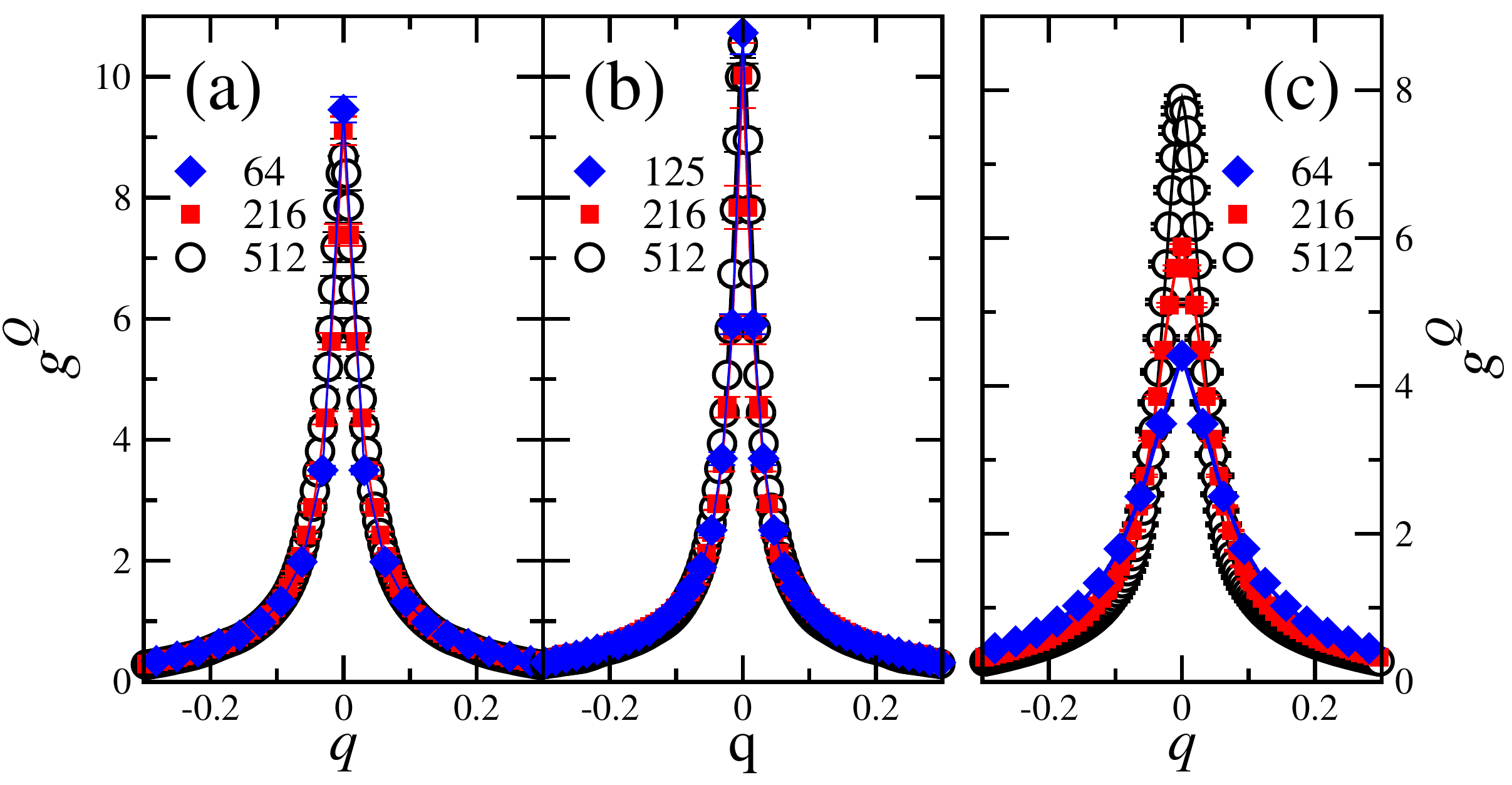}
\caption{ (Color online):
  (a) Plots of $g^Q$ vs. $q$ for  SC systems at $T=0.2$, with $Q=1/2$ and the values of $N$ indicated in the figure. (b) The same plots for RCP systems.
  (c) { The same plots for the SK model at $T=0.24$. Note that $T \approx T_{sg}/4$ in all three cases.} }
\label{gqtrio}
\end{figure}

At very low temperatures, $g^{Q}(q)$ could be interpreted as the averaged shape of all CO spikes provided that individual spikes in $p_{\cal J}(q)$ distributions do not overlap each other.\cite{PADdilu2} Temperatures as low as $T_{sg}/4$ are needed to observe this regime (see Fig.~\ref{p_delta}(a)).
Plots of $g^{Q}$ vs. $q$ for $Q=1/2$ and $T=0.2$ are shown in  
Figs.~\ref{gqtrio}(a) and \ref{gqtrio}(b) for RADs on SC and RCP arrays respectively, and in Fig.~\ref{gqtrio}(c) for the SK model at $T=0.24$. $g^{Q}$ curves appear to be very pointed and narrow in all cases. 
Note that curves for RAD systems in Figs.~\ref{gqtrio}(a) and \ref{gqtrio}(b) do not change appreciably with
system size for both SC and RCP arrangements. In contrast, $g^{Q}$ curves for SK in  Fig.~\ref{gqtrio}(c) become sharper as $N$ increases. This is as expected for the mean-field RSB scenario, in which CO spikes 
become Dirac delta functions in the macroscopic limit.

\begin{figure}[!b]
\includegraphics*[width=89mm]{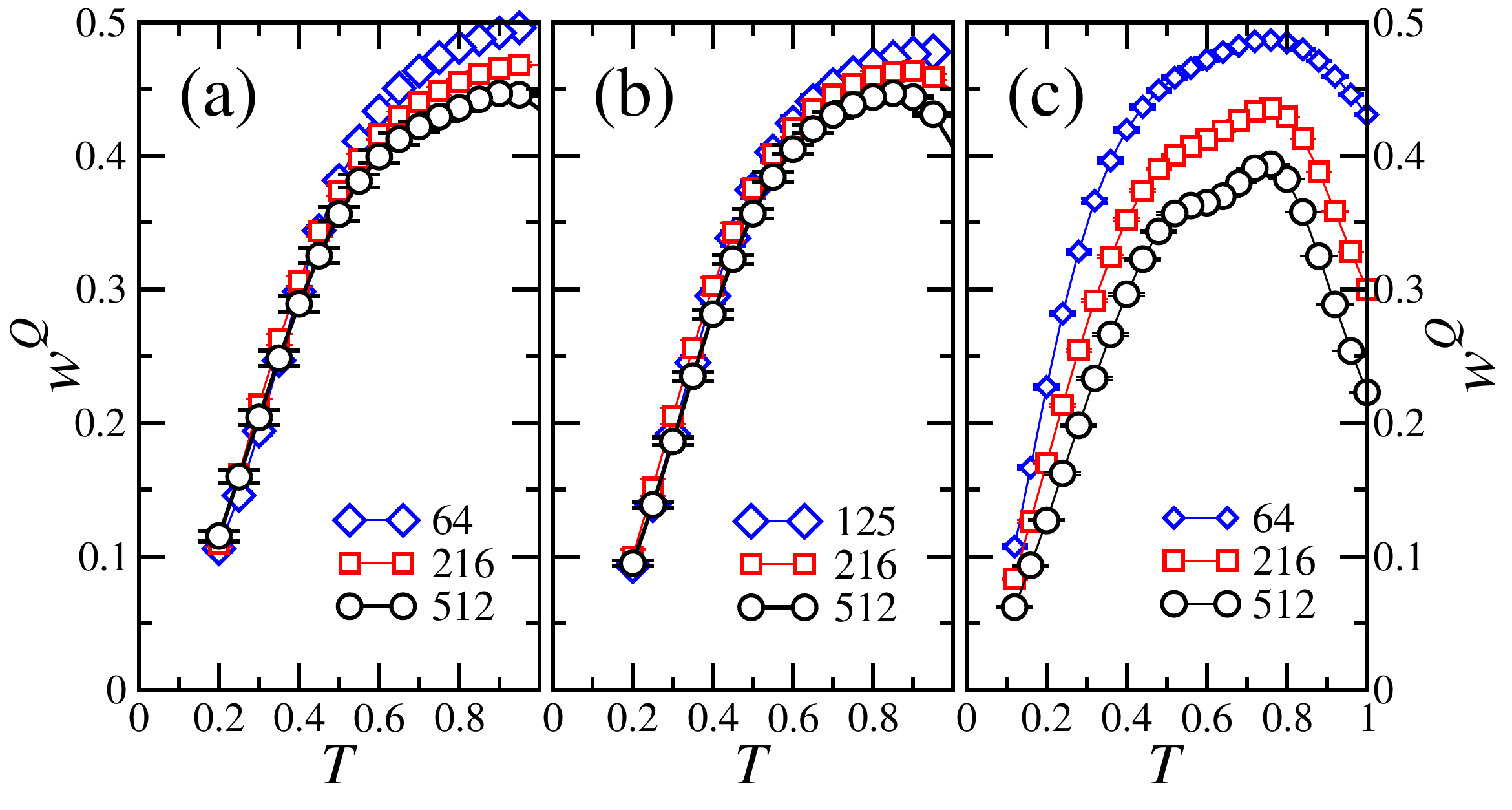}
\caption{ (Color online)
(a) Plots of $w^Q$ vs. $q$ for SC systems for $Q=1/2$ and the values of $N$ indicated in the figure. 
(b) The same plots for RCP systems. (c) The same plots for the SK model.}
\label{widths}
\end{figure}

Given that $g^{Q}(q)$ is a normalized distribution, we compute its width as 
$w^{Q} \equiv1/g^{Q}(0)$.\cite{pcf}
Plots of $w^{Q}$ vs. $T$ are shown in   Figs.~\ref{widths}(a) and \ref{widths}(b)
for RADs on SC and RCP arrays respectively. In both cases, $w^{Q}$ does not change appreciably
with system size, suggesting that the width of CO spikes do not vanish in the $N\rightarrow\infty$ limit for
all temperatures below $T_{sg}$.
On the other hand, plots of $w^{Q}$ vs. $T$  for the SK model displayed in Fig.~\ref{widths}(c) 
indicate a vanishing width (curves actually go like $w^Q\sim N^{-3/2}$), \cite{PADdilu2} in agreement with the RSB picture.

\subsection{Cumulative Distributions}
\label{cumul}

\begin{figure}[!t]
\includegraphics*[width=80mm]{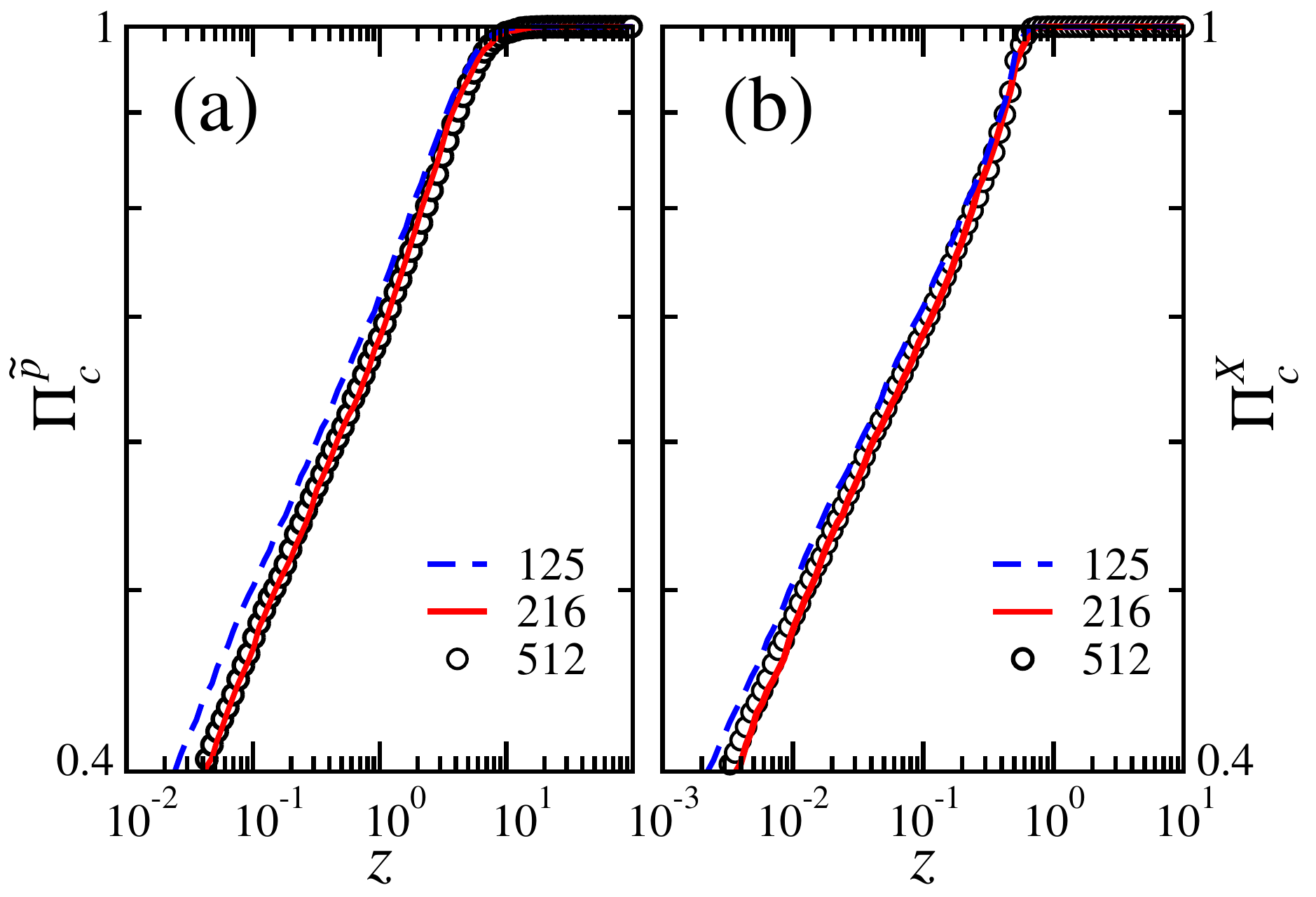}
\caption{ 
 (Color online) (a) Plots of the cumulative distribution $\Pi_c^{\widetilde{p}}$ vs. $z$ 
for RCP systems, with $Q=1/2$, $T=0.2$ and the values of  $N$ indicated in the figure.
 { (b) Same as in (a) for the  cumulative distribution $\Pi_c^{X}$.}
 } 
\label{piequis}
\end{figure}

Pair correlation functions do not provide information about the height of CO spikes in the  $(-Q,Q)$ region. Yucesoy  {\it et al.}\cite{yucesoy}
have proposed an observable that depends on the height of CO spikes in SG models. 
They consider the maximum value of $p_{\cal J}(q)$ for $q \in (-Q,Q)$,
\begin{equation} 
\\\widetilde{p}_{\cal J}^{~Q} \equiv\max\{~p_{\cal J}^{Q}(q) : |q|<Q~\},
\label{peaksq}
\end{equation}
and count a sample as \emph{peaked} if $\widetilde{p}_{\cal J}^{~Q}$ exceeds some specified value $z$.
Then, the cumulative distribution $\Pi_{c}^{\widetilde p} (z)$ of \emph{non-peaked} samples 
is computed as a function of $z$. Plots of $\Pi_{c}^{\widetilde p}$ vs. $z$ for RAD on RCP arrays  are shown in Fig.~\ref{piequis}(a) for $T=0.2$ and $Q=1/2$. They suggest that  $\Pi_{c}^{\widetilde p}$ becomes size-independent for  $N\ge216$. In contrast, simulations for the SK model have shown that $\Pi_{c}^{\widetilde p}(z)$ clearly decreases as $N$ increases for $z\gtrsim 0.5$ at low temperatures.\cite{PADdilu2}
The latter is in agreement with the RSB picture, for which CO spikes become Dirac delta functions in the $N\to\infty$ limit. 

Given that $X_{\cal J}^{Q}$ is a ($\cal J$-dependent) random variable, it is interesting to explore how this variable is distributed.
Following Ref.~\onlinecite{billoire}, we  define its cumulative distribution $\Pi_{c}^{X} (z)$ as the fraction of samples having $X_{\cal J}^{Q} < z$.
Semilog plots of  $\Pi_{c}^{\it X}$ vs. $z$ for RAD on RCP arrays displayed in Fig.~\ref{piequis}(b) appear to be size independent and exhibit a power-law behavior of $\Pi_{c}^{\it X}(z)$ for small $z$. 
Our results for $\Pi_{c}^{\it X}$  are not in contradiction with a RSB scenario. 

\section{CONCLUSIONS AND DISCUSSION}
\label{conclusion}

We have studied ensembles of dense packings of identical classical Ising dipoles at low temperature. Each dipole is the total magnetic moment of a
single-domain spherical NP (due to their smallness, NP admit one single domain). { We assume that the local anisotropies of the NP oblige the magnetic
moment to lie along an easy axis.} The arbitrary orientation of all NP provokes that each dipole is randomly oriented.

We consider random axis dipole (RAD) systems with two types of packings: (i) the NP are placed on SC lattices and (ii) random closed
packings (RCP) that fill a volume fraction $\phi_{0}=0.64$, as in recent experiments with maghemite NP.

We have focused on the role played by the dipolar interaction in the thermodynamical equilibrium.
Dipoles were allowed to flip between up and down directions along one of their easy axes, regardless of the height of the anisotropy barriers.

Previous simulations for RADs on SC lattices provided evidence for the existence of a transition from paramagnetic to SG phases.
However, numerical results did not lead to any firm conclusion on the nature of this SG phase. From the study of the overlap parameter $q$ and their
associated correlation length, we have found a marginal behavior for RADs on SC and RCP lattices for
temperatures below a temperature $T_{sg}$. 
Actually, the similarities between SC and RCP systems extend to all observables we have explored.

From the variation of $\xi_L$ with $T$ and $L$ we have found  $T_{sg}=0.75(2)$  and $T_{sg}=0.78(3)$ for SC and RCP lattices respectively.
In the SG phase we have observed (i) an algebraic decay of $q_2$ with the system size, and (ii) absence of any divergence in
the values of the ratio $\xi_L/L$ as the system size increased (see section III.A). 
This marginal behavior fits neither the droplet model nor a RSB scenario.  

In spite of the existence of \emph{quasi-long-range} SG order, the overlap distributions $p_{\cal J}(q)$ are comb-like and markedly sample dependent, 
like it occurs for the EA and the SK models.\cite{yucesoy,pcf,aspelmeier} We have studied the sample-to-sample statistics of $p_{\cal J}(q)$ for
$q\in(-Q,Q)$ at low temperatures (section III.B), finding
that $p(q)$ and $\delta p(q)^2$ (as well as $X^{Q}$ and $\Delta ^{Q}$) do not vary  with $N$. By computing also
the averaged width $w^{Q}$ of the spikes found in $p_{\cal J}(q)$ distributions (section III.C), we conclude that
$w^{Q}$ does not vanish in the thermodynamic limit. Accordingly, the fraction of samples with spikes higher than
a certain threshold does not change with $N$ at low temperatures (section III.D). 

Altogether, these results are in agreement with the above-mentioned marginal behavior.
Our MC data for the SK model illustrate that there is some conflict between RSB predictions and our results for RAD. It is worth mentioning that
{our findings for densely packed RAD systems resemble the behavior of RAD
systems} with strong dilution\cite{RADdilu} and that of systems of parallel Ising dipoles in site-diluted lattices\cite{PADdilu,PADdilu2}.




\section*{Acknowledgements}
We thank the Centro de Supercomputaci\'on y Bioinform\'atica  at University of M\'alaga, and to 
Institute Carlos I at University of Granada  for their generous allocations of computer time in clusters Picasso, and Proteus.
J.J.A. thanks for financial support from the Ministerio de Econom\'ia y Competitividad of Spain,
through Grant No. FIS2013-43201-P. We are grateful to J. F. Fern\'andez for helpful discussions. 


\begin{thebibliography}{99}
\bibitem{remark} One could argue, however, that larger values of $N$
are needed to discard a scenario with a non-vanishing $q_{2}$ and a diverging correlation length.
\bibitem{remark2} This might seem to be in contradiction to the fact that $\xi_L/L$ curves do clearly cross, as shown in
  Figs. \ref{longsBIS}(a) and \ref{longsBIS}(b), and that, as pointed out in Ref. \onlinecite{balle},
$\xi_L/L$ vs. $T$ curves merge, not cross, for the 2D XY model, as $T\to T_{sg}$ from above.
Some specific examples in Ref.~\onlinecite{PADdilu} illustrate how both merging and spreading  
as $T$ decreases below $T_{sg}$ can occur depending on some minor details of  $G(r)$.
\end{thebibliography}
\end{document}